\documentclass[11pt]{article}

\usepackage[T1]{fontenc}
\usepackage[utf8]{inputenc}
\usepackage{lmodern}
\usepackage[margin=1in]{geometry}
\usepackage{microtype}
\usepackage{amssymb}
\usepackage{array}
\usepackage{longtable}
\usepackage{booktabs}
\usepackage{ragged2e}
\usepackage[most]{tcolorbox}
\usepackage{authblk}          

\usepackage[hidelinks]{hyperref}
\usepackage{cite}

\newcolumntype{P}[1]{>{\RaggedRight\arraybackslash}p{#1}}

\newcommand{\chk}[1]{\par\noindent\hangindent=1.4em\hangafter=1\makebox[1.4em][l]{$\square$}#1\par}

\newcommand{\protocol}[2]{\par\medskip\noindent\textbf{Protocol #1:} \textit{#2}\par}

\title{Can we create a `race to the top' for weather forecasts to inform smallholder farmer decisions?}

\author[1]{Colin Aitken\thanks{Corresponding author: \texttt{caitken@uchicago.edu}}}
\author[2]{Michael K. Tippett}

\author[3,4]{Pedram Hassanzadeh}
\author[4]{Katherine Kowal}
\author[5,6]{Rendani Mbuvha}
\author[7]{John H. Marsham}
\author[6,8]{Shruti Nath}
\author[9]{Ousmane Ndiaye}
\author[7,10]{Douglas J. Parker}
\author[7]{Caroline M Wainwright}

\author[1,11,12]{Michael Kremer}
\author[13,14]{William R. Boos}

\affil[1]{Development Innovation Lab, University of Chicago, Chicago, USA}
\affil[2]{Department of Applied Physics and Applied Mathematics, Columbia University, New York, USA}
\affil[3]{Department of the Geophysical Sciences, University of Chicago, Chicago, USA}
\affil[4]{Data Science Institute, University of Chicago, Chicago, USA}
\affil[5]{School of Statistics and Actuarial Science,
University of Witwatersrand, South Africa}
\affil[6]{AfriClimate AI, South Africa}
\affil[7]{School of Earth, Environment and Sustainability, University of Leeds, Leeds, United Kingdom}
\affil[8]{Department of Physics,
University of Oxford, Oxford, United Kingdom}
\affil[9]{African Centre of Meteorological Applications for Development, Niamey, Niger}
\affil[10]{NORCE Research AS Climate Risk, Bergen, Norway}
\affil[11]{Kenneth C. Griffin Department of Economics, University of Chicago, Chicago, USA}
\affil[12]{Harris School of Public Policy, University of Chicago, Chicago, USA}
\affil[13]{Climate and Ecosystem Sciences Division, Lawrence Berkeley National Laboratory, Berkeley, USA}
\affil[14]{Department of Earth and Planetary Science, University of California, Berkeley, USA}

\date{October 7, 2026}

\begin{document}

\maketitle
\begin{abstract}
Artificial-intelligence weather prediction (AIWP) models have made it possible to produce high-quality tailored forecasts with limited computational resources. This advance has the potential to benefit hundreds of millions of farmers in low- and middle-income countries who lack access to forecasts of critical weather phenomena. However, it can be difficult for key stakeholders to evaluate forecast quality, risking a ``race to the bottom'' as cheap but low-quality forecasts crowd out forecasts that would benefit farmers. We propose a set of principles and protocols for evaluating agriculturally-relevant forecasts as a starting point for standards that would let forecasters credibly convey their forecasts' quality.
\end{abstract}

\section{Introduction and Overview}

Farmers benefit from access to weather forecasts, which allow them to make more informed decisions about the choice of inputs, the timing of planting and harvesting, and other key agricultural practices~\cite{burlig2024_value_of_forecasts,fosu2018disseminating, yegbemey2021_sms_maize, rudder2024learning, gine2017forecasting}. Recent programs have shown that agriculturally-relevant forecasts can be delivered cheaply to millions of farmers\cite{pib2026aiagri}, and governments and funding agencies are increasingly interested in forecast dissemination as a means of improving both food production and farmer livelihoods.

The development of new artificial-intelligence weather prediction (AIWP) models provides an opportunity to further expand forecast access. Open-source AIWP models meet or exceed the skill of leading numerical weather prediction (NWP) models on many metrics while requiring a fraction of the computational resources~\cite{allen2025end, ben2024rise,bi2023accurate,lam2022graphcast,pathak2022fourcastnet,price2025probabilistic, nath2026predicting}. This reduced computational cost lowers the barrier to entry: tailoring a state-of-the-art forecast for a particular purpose no longer requires a national meteorological center's supercomputing budget, and so startups, academic groups, app developers, and individual physical and computational scientists are beginning to produce agriculturally-focused forecasts. This expansion in forecast production includes both high- and low-quality forecasts, and it can be challenging to distinguish between the two.

When farmers use weather forecasts to inform high-stakes decisions, accurate forecasts provide substantial value. For the same reason, inaccurate forecasts can cause substantial harm. For example, farmers who plant because they were (incorrectly) assured that continuous rains had started risk losing their crop if an unexpected dry spell follows. Moreover, the introduction of low-quality forecasts can depress future demand even for subsequently improved products: farmers who have been harmed by inaccurate forecasts are less likely to trust future promises of quality, and their governments may be less likely to disseminate them. 

We illustrate some of the challenges in identifying good forecasts through two examples.  A forecast distributor may assume that off-the-shelf models developed by leading tech companies or international forecasting centers are high-quality and therefore safe to rely on. In practice, however, forecast quality is multidimensional: a global model may skillfully predict some quantities, but may have biases or failure modes that render its deployment inappropriate for a particular use case (Illustrative Example 1, see also~\cite{murphy1993good}). At the same time, verifying a forecast for a particular use case is technically demanding, and seemingly minor methodological errors can lead even highly unreliable forecasts to appear accurate (Illustrative Example 2). 

AIWP models themselves raise new challenges for verification\cite{nath2026can}.  The fact that many of these models are trained on nearly all of the available weather data creates a risk that they are overly constrained by historical events and unable to predict out-of-sample future events (i.e., they are overfitted).  Using observational networks in verification that were not explicitly employed in model training can provide false reassurance if those data come from the same time periods used in training, because the networks measure the same atmospheric states on which the models were trained.  The computational efficiency of AIWP models also make it easy to produce `hyper-local' or `farm-level' forecasts that appear realistic to a casual viewer, yet it may be impossible to obtain the high-resolution observational data needed to verify such forecasts, or to guarantee that such data were not used in model training.

The difficulty of assessing forecast quality can lead to a ``race to the bottom''. If groups procuring forecasts, such as large funders and government agricultural agencies, cannot reliably distinguish good from bad forecasts, then forecast providers will have a financial incentive to save money even at the cost of reducing forecast quality. If left unchecked, low-quality forecasts can fully crowd out the kind of forecasts that would benefit farmers, a classic pattern of market failure induced by information asymmetry that economists have studied for decades~\cite{akerlof1970market}.  

To instead create a ``race to the top'', forecast producers need a way to accurately convey the quality of their forecasts. This process needs to be simple enough that a small team can implement it on a realistic timeline and an evaluator without deep methodological training can confirm it, while also being comprehensive enough that a stakeholder relying on them can be confident in a forecast's quality. We propose that a group of researchers, nonprofits, companies, meteorological agencies, and others could agree to a set of principles and protocols to be used when designing and evaluating forecasts intended for large-scale agricultural dissemination. (Similar principles will likely prove useful for other kinds of dissemination, such as for health purposes, but we focus here on agriculture for concreteness.) Such a group would need a process for training new groups on the principles, certifying that the principles had been followed, and building consensus both on an initial set of principles and a process for amending them as new issues arise and the underlying science evolves.  These principles could then be further iterated on and institutionalized by groups like the World Meteorological Organization or regional agencies. 

A clear set of principles for certifying quality would benefit both producers and consumers of forecasts: groups interested in dissemination could then select high-quality forecasts that would benefit the farmers who receive them and build trust to incorporate forecasts into a wider range of decisions. Forecast coverage would expand, as funders and governments who were previously overwhelmed by the number of available products or deterred by the risks of low-quality forecasts would have a new tool to ensure a baseline level of reliability.  Producers of forecasts would be incentivized to produce high-quality forecasts, and would themselves benefit from knowing ahead of time what rules they need to follow and what information they will need to provide, rather than responding to idiosyncratic requests after investing time and money into a particular model.

In this paper, we propose a set of principles and protocols we believe would provide a baseline level of quality assurance for agriculturally-relevant weather forecasts, collecting best practices for evaluation from meteorology, social science, statistics, and computer science. We focus primarily on verifying the forecasts themselves; in Principle 4 we sketch some considerations related to the quality of the dissemination process, but we do not develop these as deeply as the forecast verification principles.   Table~\ref{tab:short} gives a list of principles, along with an example protocol (and failure of protocol) for each. We provide a full ``checklist'' of protocols in Supplementary Table~\ref{tab:full}, and a fuller description of each in a supplemental appendix. 

We see this as a starting point for iteration as we work to build consensus across a wider group of stakeholders, including forecast developers, scientists, meteorological and agricultural agencies,  funders, and others. We welcome any comments and ideas for collaboration. 
\clearpage

\begin{tcolorbox}[enhanced, breakable, colback=gray!6, colframe=gray!60, fonttitle=\bfseries, title=Illustrative Example 1]
Some farmers use rainfall forecasts to plan harvesting activities ahead of time, since harvesting on a rainy day risks spoiling crops. An agricultural minister arranges to disseminate forecasts from a leading global forecasting model to help farmers with this decision, after checking that this forecast shows high skill on standard metrics including Root Mean Squared Error (RMSE) for rainfall in the minister's country.

Unbeknownst to the minister, this model has a known ``drizzle'' bias: it frequently shows small amounts of rainfall on days that turn out to be completely dry. (This problem is observed in several top AIWP models~\cite{moldovan2026aifs}.) This does not affect the overall forecast RMSE very much, since the model is good at distinguishing between light, medium, and heavy rain. However, it makes the raw model outputs misleading when used for this particular decision,  predicting more than twice as many wet days as actually occur and leaving farmers with few usable harvest windows.
\end{tcolorbox}

\begin{tcolorbox}[enhanced, breakable, colback=gray!6, colframe=gray!60, fonttitle=\bfseries, title=Illustrative Example 2]
A group claims to be able to predict when large-scale floods will occur. They show that their model predicted the starting date of the last 3 large-scale floods to within 1 day of the true beginning. An agency invests heavily in the flood forecasts, and prepares a major response when the model predicts a large flood. No flood occurs: it turns out the model has in fact predicted ``10 of the last 3 large-scale floods'', and their evaluation never reported the false-alarm rate\footnote{Paul Samuelson, responding to claims that Wall Street indexes had predicted four of the last five recessions, famously wrote that they had in fact predicted ``nine out of the last five recessions''~\cite{samuelson1966science}.}. In addition to wasted money, this public failure reduces trust in weather forecasts across the board.
\end{tcolorbox}

{\small
\begin{longtable}{P{0.31\textwidth} P{0.26\textwidth} P{0.35\textwidth}}
\caption{Principles and Example Protocols}\label{tab:short}\\
\toprule
\textbf{Principle} & \textbf{Example Protocol} & \textbf{Example Protocol Failure} \\
\midrule
\endfirsthead
\toprule
\textbf{Principle} & \textbf{Example Protocol} & \textbf{Example Protocol Failure} \\
\midrule
\endhead
\bottomrule
\endfoot

\textbf{Transparency:} Claims of forecasting skill should be supported with enough information for an independent evaluator to understand, assess, and if necessary reproduce the claim.
&
Evaluation methodology is described clearly enough that reviewers can replicate the evaluation. For any unclear points, applicants are responsive to questions.
&
Groups A and B bring forecasts, claiming improvements of 5\% and 30\% over a climatological baseline, respectively. Because the groups do not provide information on methodology, the disseminator does not realize group A compared their forecasts out-of-sample to a very strong baseline, for which a 5\% improvement requires a very good model, while group B compared the model to a poorly constructed climatological baseline on the same data their model was trained on, so that a $30\%$ improvement reflects a very low-quality model.  Without this knowledge, the disseminator wrongly chooses group B's model and farmers receive highly inaccurate weather information.
\\ \midrule

\textbf{Relevant Evaluation:} An evaluation should provide information about a forecasting system's strengths and weaknesses relative to other available options, including climatological information and open-source alternatives. The evaluation should be thorough and avoid cherry-picking skill metrics that make the model look good. Uncertainty should be quantified when possible to help readers distinguish genuine differences from noise.
&
Forecast compared to decision-relevant baselines, including climatology or persistence and leading open-source models.
&
A startup shows that a forecast model can predict short-term rainfall 30\% better than a climatological baseline. An impressed government official pays the startup millions of dollars for the rights to disseminate their forecasts, not knowing that typical open source forecasting models are easy to run and would have been able to predict short-term rainfall 45\% better than the same climatological baseline. The official has therefore paid for a product that is worse than the standard freely available forecasts many meteorologists rely on.
\\ \midrule

\textbf{Demonstrated Out-of-Sample Skill:} It is often much harder to predict future weather than to predict past realized weather in forecasting exercises (e.g. hindcasts). To avoid data leakage, forecast skill should be evaluated on a ``test set'' consisting of data the model was not explicitly trained on. However, the risk of leakage is not limited to explicit model training: decisions about model architecture, features, thresholds, and regions made on the basis of test set performance can also inflate reported forecast skill. Forecasters should take steps to avoid this, and explicitly communicate these steps.
&
Precise description of how training and test data were separated (e.g.\ using cross-validation or holding out a specific test set) provided, and what forms of model selection or fine-tuning occurred on test data.
&
A group shows that their model, trained on data from 2000-2020, is able to predict extreme rain events in their location in 2021-2025, which they describe as a held-out test set. A funder invests in disseminating these forecasts, but operational performance is much worse than their evaluation indicated. Upon further investigation, it turns out the group made a number of modelling decisions (including which features to use for prediction) based on performance on the 2021-2025 test set. As a result, this was no longer a clean test set, and their metrics overestimated performance.
\\ \midrule

\textbf{Decision-Relevance:} A forecast needs to be useful to (and understandable by) its intended audience.
&
User comprehension and actionability tested (e.g.\ through A/B tests and focus groups)
&
A government agriculture agency procures highly-skilled rain forecasts, and disseminates forecasts telling farmers there is a 90\% chance of rain in the next week. Farmers, however, interpret this to mean there will be a large quantity of rain, rather than a high chance of (potentially very light) rain, and take costly and unnecessary actions to avoid flooding.
\\
\end{longtable}
}


\clearpage
\appendix
\renewcommand{\thetable}{S\arabic{table}}
\setcounter{table}{0}

\section*{Supplementary Information}
\addcontentsline{toc}{section}{Supplementary Information}

{\small
\begin{longtable}{P{0.44\textwidth} P{0.50\textwidth}}
\caption{Full Table of Principles and Protocols}\label{tab:full}\\
\toprule
\textbf{Principles} & \textbf{Protocols} \\
\midrule
\endfirsthead
\toprule
\textbf{Principles} & \textbf{Protocols} \\
\midrule
\endhead
\bottomrule
\endfoot

\textbf{Transparency:} Claims of forecasting skill should be supported with enough information for an independent evaluator to understand, assess, and if necessary reproduce the claim.
&
\chk{Evaluation methodology is described clearly enough that reviewers can replicate the evaluation. For any unclear points, applicants are responsive to questions.}

\smallskip
\chk{Humans able to certify principles were followed.}

\smallskip
\chk{Evaluation data and code available on request or (preferably) made public}

\smallskip
\chk{Hindcast-vs-operational differences in forecast systems are disclosed, with expected model version changes (e.g. NWP models commonly change versions annually)}

\smallskip
\chk{Text of the actual messages to be distributed to farmers provided for review}

\smallskip
\chk{Archiving, registration, or escrow performed for operational forecasts, to preserve trust}
\\ \midrule

\textbf{Relevant Evaluation:} An evaluation should provide information about a forecasting system's strengths and weaknesses relative to other available options, including climatological information and open-source alternatives. The evaluation should be thorough, and avoid cherry-picking skill metrics that make the model look good. Uncertainty should be quantified when possible to help readers distinguish genuine differences from noise.
&
\chk{Forecast compared to decision-relevant baselines, including climatology or persistence and leading open-source models.}

\smallskip
\chk{Comparison is made for the actual phenomenon of interest (e.g.\ RMSE for precipitation levels is not a test of whether a model captures when a dry season will transition into a wet season)}

\smallskip
\chk{Ground-truth dataset named; regional biases discussed; robustness across datasets disclosed where feasible/relevant.}

\smallskip
\chk{Standard metrics included for the phenomenon of interest (e.g.\ Brier Score, AUC, RPS or CRPS for probabilistic forecasts, contingency tables for deterministic forecasts)}

\smallskip
\chk{Forecasts scored under a realistic operational cadence (e.g.\ weekly forecasts, rather than forecasts initialized one week ahead of a known event) to identify false alarms and forecast misses}

\smallskip
\chk{Probabilistic forecasts checked for calibration (e.g. do events forecast with 30\% probability end up occurring 30\% of the time?)}

\smallskip
\chk{Uncertainty quantified; spatial differences in forecast skill are considered only if they are larger than a plausible uncertainty}
\\ \midrule

\textbf{Demonstrated Out-of-Sample Skill:} It is often much harder to predict future weather than to predict past realized weather in forecasting exercises (e.g. hindcasts). To avoid data leakage, forecast skill should be evaluated on a ``test set'' consisting of data the model was not explicitly trained on. However, the risk of leakage is not limited to explicit model training: decisions about model architecture, features, thresholds, and regions made on the basis of test set performance can also inflate reported forecast skill. Forecasters should take steps to avoid this, and explicitly communicate these steps.
&
\chk{Precise description of how training and test data were separated (e.g.\ using cross-validation or holding out a specific test set) provided, and what forms of model selection or fine-tuning occurred on test data.}

\smallskip
\chk{Post-hoc changes to the phenomenon definition (including thresholds, forecasting regions, lead times etc.) should be clearly communicated, and all metrics should be reported with both the original definition and the new definition for comparison.}
\\ \midrule

\textbf{Decision-Relevance:} A forecast needs to be useful to (and understandable by) its intended audience.
&
\chk{Decision-relevance articulated, naming a specific user decision at the delivered lead time}

\smallskip
\chk{User comprehension and actionability tested (e.g. through A/B tests and focus groups)}

\smallskip
\chk{When feasible, sensitivity of the predicted metric(s) to any underlying thresholds or parameters examined, and established to be sufficiently low for usability}
\\
\end{longtable}
}

\clearpage
\section*{Details on protocols}

\subsection*{Principle 1: Transparency}

Claims of forecasting skill should be supported with enough information for an independent evaluator to understand, assess, and if necessary reproduce the claim.

\protocol{1a}{Evaluation methodology is described clearly enough that reviewers can replicate the evaluation. For any unclear points, applicants are responsive to questions.}

Estimates of forecast skill are only meaningful in the context of how the estimates were produced. For example, the mean absolute error (MAE) of a monsoon onset date forecast initialized a few days before the true onset in a location might be quite low compared to the MAE of an onset forecast initialized on a fixed calendar date without reference to the true onset date. The same raw MAE estimate could reflect a very bad model in the former case, and a very good model in the latter case, and a reviewer without full understanding of how each metric was produced will not be able to determine which is the case.

More generally, rigorous evaluation methodologies tend to produce more realistic, and hence more moderate, estimates of forecast skill. If reviewers do not have access to the methodology used to support a claim of forecast skill, then applicants will not have an incentive to use the best available methodology, and may instead use methods that make their forecasts appear much better than they actually are. This creates an adverse selection problem, directing funding towards groups with worse methodological practices (which may mean lower-quality forecasts).

\protocol{1b}{Humans able to certify principles were followed.}

As Artificial Intelligence (AI) capabilities increase, researchers and developers are able to hand off more tasks to automated AI models. Frontier models under certain conditions engage in ``reward hacking'', using unintended methods to fool a grader into believing a task has been completed (e.g. by hacking an internal system to find an answer key).

Evaluating weather models, particularly for rare events, depends on clear distinctions between training and validation datasets and test datasets (see Principle 3, below). For example, it is important that a researcher be able to confirm that test set data was not used to train or choose features for a forecasting model. If the model is designed by AI agents, the researcher needs some way to confirm that the AI agents did not cheat by using test set data in inappropriate ways. The scientific community typically uses trust and reputation to ensure researchers tell the truth about how their datasets were used. However, these mechanisms break down if AI models are used in a way where human researchers or developers do not know if the principles were followed, because they do not have sufficient insight into what their AI agents are doing.

As AI models evolve, the ways humans interact with them will as well. We do not take a stance on what this certification should look like (for example, humans might restrict the use of AI agents to individual, well-specified tasks to ensure they have full control over all research decisions, or future AI models might be designed in a way such that it is possible to validate precisely what information they used to make which decisions).

\protocol{1c}{Evaluation data and code available on request or (preferably) made public}

Evaluation code can clarify methodological choices that are not clear from prose descriptions; it can also contain coding errors that change the results of the analysis. Access to this code allows independent audits, replication, and expansion on a given finding. Data and code used in benchmarking should be made available to key stakeholders (with a reasonable amount of advance notice, ideally specified ahead of time). More weight should be given to benchmarking results when these are made public and available for peer review.

\protocol{1d}{Hindcast-vs-operational differences in forecast systems are disclosed, with expected model version changes (e.g. NWP models commonly change versions annually)}

Operational models can perform worse than the hindcasts used for benchmarking; for example, models initialized on real-time initial conditions may perform differently from hindcasts initialized on atmospheric reanalyses. Forecasters should be transparent about any differences, and how they would expect those differences to affect forecasts. When possible, skill scores should be tested for robustness to these differences. In some cases, the model itself may change throughout the hindcast period, particularly for Numerical Weather Prediction (NWP) models where producing a large series of hindcasts may be computationally expensive. When feasible, the same model should be used for hindcasts and operations. Similarly, any changes to the model between past verification data and a current version should be transparently communicated.

\protocol{1e}{Text of the actual messages to be distributed to farmers provided for review}

The actual message distributed to a farmer may differ from the numbers produced by an underlying forecast model: for example, a forecasting model might estimate the probability that the rainy season will start on each of the next 30 days, but farmers might receive the median of this distribution, the probability that onset occurs in the next week, or agricultural advice estimated from these probabilities. Simplifying information this way can increase comprehension, but can introduce two problems that can be identified from the message text itself. First, a model may show skill when its full, probabilistic output is tested, but have little or no predictive power when this is collapsed to a single number. In this case, the summary number should be tested as a forecast in its own right (see Protocols 2b, 2e). Second, the information provided may require further support: the forecasting skill of the underlying model should not be taken as evidence that the agricultural advice based on it is good; evidence should be provided that the advice given is better than what a farmer (who has much more information about their personal circumstances and land) would do themselves with the weather information.

\protocol{1f}{Archiving, registration, or escrow performed for operational forecasts, to preserve trust}

Once operational forecasts begin, claims of forecasting skill are most reliable if registered in real-time on a platform (e.g.\ Zenodo) where their skill can be independently verified after-the-fact. This is most credible when done at the time of the forecast, not at the end of the season after the realized weather outcomes have been observed. For-profit companies, who may not be willing to do this in real-time, can find a trusted, independent third-party who can hold their forecasts ``in escrow'' for future verification, or publish an encrypted or hash version of the forecast that a later plaintext version can be checked against.

\subsection*{Principle 2: Relevant Evaluation}

An evaluation should provide information about a forecasting system's strengths and weaknesses relative to other available options, including climatological information and open-source alternatives. The evaluation should be thorough, and avoid cherry-picking skill metrics that make the model look good. Uncertainty should be quantified when possible to help readers distinguish genuine differences from noise.

\protocol{2a}{Forecast compared to decision-relevant baselines, including climatology or persistence and leading open-source models.}

A forecasting model intended for dissemination must be benchmarked against a reasonable baseline to determine whether or not it shows sufficient accuracy/skill. The definition of a ``reasonable'' baseline will be phenomenon-dependent and may change over time as different information becomes publicly available. For short- and medium-to-long range forecasts, persistence (using very recent weather) or climatology (using the historical distribution of weather) will often be reasonable baselines. Climatology may need to be adjusted to account for temporal trends. Climatological baselines can inappropriately inflate apparent forecast skill if they aggregate at the wrong level~\cite{hamill2006measuring} or fail to account for decision-relevant information included in the benchmark design~\cite{aitken2026designing}. In the dissemination context, the most decision-relevant baseline may also include the best forecast information already freely available to the intended users, e.g. forecasts provided by the national meteorological service, when the data required to benchmark these is available.

The profit potential from proprietary forecasting models can incentivize startups and existing companies to invest in weather forecast innovation. However, proprietary models also have drawbacks: independent evaluation is more difficult, the field as a whole may not benefit from discoveries that are not disclosed, and meteorological agencies may not be willing to rely on the ongoing goodwill of a private actor. As a result, a group pitching a proprietary model carries a burden of proof to show that their model is better than leading open-source models for a particular task: this keeps the incentive to innovate when this innovation improves on the state-of-the-art, but prevents private actors from charging rents for information that is lower-quality than what's already freely available. Ideally, the open-source models should be bias-corrected and calibrated for the particular use case and (when possible) blended~\cite{aitken2026designing} to ensure a fair comparison. 

\protocol{2b}{Comparison is made for the actual phenomenon of interest (e.g.\ RMSE for precipitation levels is not a test of whether a model captures when a dry season will transition into a wet season)}

The baseline comparison must be relevant to the actual information being disseminated: for example, if a forecaster would like to issue information about potential dry periods, the metrics should specifically reflect the model's ability to predict dry periods, not merely general metrics related to rainfall quantity. (See Illustrative Example 1 in the main text for an example of how these might differ).

\protocol{2c}{Ground-truth dataset named; regional biases discussed; robustness across datasets disclosed where feasible/relevant.}

In many regions where mass dissemination is most valuable, the ``truth'' itself is uncertain: rain gauges are sparse, and satellite, blended, and reanalysis products can disagree on the phenomena of interest (e.g. whether a dry spell occurred). Forecasters must state which ground-truth dataset is used and why, discuss its known biases in the region, and, where feasible, show that claimed skill is robust across plausible alternative ground truths. If skill depends materially on the choice of ground truth, this must be disclosed.

\protocol{2d}{Standard metrics included for the phenomenon of interest (e.g.\ Brier Score, AUC, RPS or CRPS for probabilistic forecasts, contingency tables for deterministic forecasts)}

Any single metric only captures a particular component of forecast performance: for example, a model might have few false positives but many false negatives or vice versa. There may often be good reason to emphasize nonstandard metrics, particularly if these are designed around a particular decision farmers might make.

To avoid the risk of cherry-picking metrics where a particular model performs well, forecast evaluations should include all standard metrics in addition to their preferred metric. For example, for deterministic binary forecasts, evaluations should include a full contingency table, including false positive and false negative rates. For probabilistic binary forecasts, Brier Scores and Area Under the Receiver Operating Characteristic Curve (AUC) should be reported, while for continuous and ordered categorical probabilistic forecasts the Continuous Ranked Probability Score and Ranked Probability Score should be reported.

In addition, basic descriptive statistics about the forecast and phenomenon should be shown. For example, for a monsoon onset date forecast it can be helpful to show the true distribution of onset dates in a location (as defined by the onset metric); for an extreme event it is helpful to show how often this event occurs, etc.

\protocol{2e}{Forecasts scored under a realistic operational cadence (e.g.\ weekly forecasts, rather than forecasts initialized one week ahead of a known event) to identify false alarms and forecast misses}

Benchmarking is most reliable when it occurs under operational conditions. This means there should be a clearly-defined period of operation in which the model is initialized at a regular cadence until either the end of the season or the occurrence of the event being forecast, whichever is relevant. Benchmarking should report scores for all forecasts in this period, not e.g.\ limiting to forecasts made a set number of days before a true event. If forecasts would only be sent under specific conditions (for example, a dry spell forecast being sent if a wet spell has first occurred), benchmarking should be done for all forecasts during the period of operation for which this condition is met. If end users receive a derived product -- for example, a binary advisory triggered by a probability threshold -- the object benchmarked should be the advisory as delivered, i.e.\ the full decision rule including issuance conditions, not only the underlying forecast field.

\protocol{2f}{Probabilistic forecasts checked for calibration (e.g. do events forecast with 30\% probability end up occurring 30\% of the time?)}

For probabilistic forecasts, the forecaster should check that both the forecast and the baseline being compared to are calibrated (in the sense that events assigned 30\% probability happen about 30\% of the time) under operational conditions. This is commonly done by showing a reliability diagram. If a climatological baseline is not calibrated, the baseline is likely either inappropriate operationally (because it does not correctly incorporate information used when deciding whether to issue a forecast or not) or the sample size is too small for the benchmark to be reliable.

\protocol{2g}{Uncertainty quantified; spatial differences in forecast skill are considered only if they are larger than a plausible uncertainty}

Any benchmarking exercise is subject to statistical noise; this is amplified for occasional events such as monsoon onset or extreme temperature or precipitation patterns. There are various ways to show this uncertainty: for example, bootstrap methods can be used to directly estimate sampling variation, or plots of benchmarking results across different years can illustrate the time-variability of the scores. If sampling variation is directly reported, appropriate methodology should be used to account for spatial and temporal autocorrelation (so that, for example, forecasts made at the same time in different locations should not be viewed as fully independent). In addition, when skill is reported across many locations, lead times, or thresholds, apparent pockets of skill can arise by chance; field-significance or false-discovery-rate methods should be used, and headline claims should not rest on the best-performing subset selected after the fact. Reviewers should take this uncertainty into account, avoiding basing decisions on comparisons that are likely to be statistical noise.

\subsection*{Principle 3: Demonstrated Out-of-Sample Skill}
It is often much harder to predict future weather than to predict past realized weather in forecasting exercises (e.g. hindcasts). To avoid data leakage, forecast skill should be evaluated on a ``test set'' consisting of data the model was not explicitly trained on. However, the risk of leakage is not limited to explicit model training: decisions about model architecture, features, thresholds, and regions made on the basis of test set performance can also inflate reported forecast skill. Forecasters should take steps to avoid this, and explicitly communicate these steps.

\protocol{3a}{Precise description of how training and test data were separated (e.g.\ using cross-validation or holding out a specific test set) provided, and what forms of model selection or fine-tuning occurred on test data.}

Benchmarking is most reliable when it is performed on out-of-sample forecasts: if a model was trained on the same data (or models are chosen based on their performance on the same data) that is being used for its evaluation, there is a high risk of overfitting, which causes benchmarking to show higher skill than the true forecast would. Similarly, if a ``test set'' is not directly used in training the model, but model features are selected by comparing how well the model performs on the test set, this introduces a risk of overfitting and should be disclosed. Any benchmarking results shared must be clear about which years are training years, which are not training years but were used for model selection, how many models were considered in the model selection procedure, etc. Tools such as cross-validation can be used when available data is sparse to increase the number of years which can be used for evaluation; note, however, that cross-validation in short climate records has known pitfalls (prominent cases have occurred where cross-validated scores were used to select among candidate models, artificially inflating the reported skill\cite{delsole2009artificial}), so cross-validated results should be accompanied by their methodology. Tests for autocorrelation can be used to determine whether the window selected for cross-validation is appropriately large. Evaluation on a training sample should be appropriately caveated and heavily discounted as evidence of forecast skill.

Furthermore, different data sources that cover the same time period and region are unlikely to contain independent information; training on one observational dataset (e.g.\ reanalysis) for a certain set of years and benchmarking against a different observational dataset (e.g.\ rain gauges) should not be viewed as using separate training/testing data. For AI models trained on global datasets, entering a new geography does not create out-of-sample data: a model trained globally through year X has effectively seen every geography through year X. Only post-training years or real-time forecasts are fully out-of-sample.

\protocol{3b}{Post-hoc changes to the phenomenon definition (including thresholds, forecasting regions, lead times etc.) should be clearly communicated, and all metrics should be reported with both the original definition and the new definition for comparison.}

Benchmarking is most reliable when the target is fixed before any evaluation occurs, so that forecasters cannot tweak the definition of the phenomenon to make their model look better. Claims of skill where a forecaster can demonstrate this (for example, because an independent group chose the target, or the forecaster publicly pre-registered the target and then evaluated forecasts of events that occurred after the pre-registration) should be given more weight. In some cases, there may be good reason to change the definition of a phenomenon: in these cases, the forecaster should disclose the changes and why they were made. Benchmarking is more reliable if the forecaster demonstrates skill on both the original and new definition, to ensure that apparent model skill is not due primarily or solely to the change. When entering a new geography, care should be taken to choose thresholds and parameters that are relevant to the intended use case; pre-registration can help avoid unintentional overfitting from trying new thresholds until skill scores are optimized.

Many social scientific fields have adopted pre-registration norms~\cite{banerjee2020praise,van2016pre} to mitigate problems where researchers can (intentionally or unintentionally) shape empirical results to fit their own beliefs~\cite{gelman2013garden}. While formal pre-registration in a public archive is rare in weather forecasting, the principle was often structurally enforced in traditional physics-based short-term forecasting by use of long-standing community metrics and real-time evaluation of new models after model development was frozen. For AI models that were trained and fine-tuned on all available data, and used to predict novel impact-oriented metrics, this may be an emerging issue.

\subsection*{Principle 4: Decision-Relevance}

A forecast needs to be useful to (and understandable by) its intended audience. Because this paper is focused on evaluating the weather forecasting systems themselves, we here flag a few relevant considerations but do not provide detailed principles or best practices for how to achieve or evaluate these goals.

\protocol{4a}{Decision-relevance articulated, naming a specific user decision at the delivered lead time}

The economic value of weather forecasts comes not merely from the importance of the phenomenon being forecast, but also the actionability of the forecast at a given lead time. For example, if a phenomenon is relevant primarily because it affects seed choice, forecasts should be provided at the time a user is purchasing seeds to be helpful.

Forecasters generally will not have full insight into how each farmer will use their forecast, or exactly how it will affect each decision. However, they should be able to articulate at least one high-value end-user decision that would be altered by potential forecasts they receive at the lead time the forecast has been tested for. Ideally, this would be supported by evidence, either from agronomic research, past experiments, or communication with farmers (e.g.\ through focus groups). Principles for evaluating this evidence are beyond the scope of this document.

\protocol{4b}{User comprehension and actionability tested (e.g.\ through A/B tests and focus groups)}

Disseminating even the best weather forecast is not useful if users cannot understand or act on the message. For example, if a user interprets a high probability of rain as a statement that there will be a large quantity of rain\cite{gigerenzer200530}, the forecasts may not benefit their decision-making and may even prove harmful. Groups interested in disseminating messages should test or even co-design their message texts with actual users to ensure messages are comprehensible and actionable by their target audience.

\protocol{4c}{When feasible, sensitivity of the predicted metric(s) to any underlying thresholds or parameters examined, and established to be sufficiently low for usability}

Many forecast targets depend on specific thresholds and parameters: for example, a ``dry spell'' might require rainfall to be under some threshold for a set number of days (e.g.\ 10 days with less than 5 mm total rain). These quantitative definitions are valuable for designing and evaluating models, but the thresholds themselves are at least somewhat arbitrary and may not be obvious to users. For example, a model that correctly predicts a 0\% chance of a dry spell because the relevant 10-day period had 5.1 mm of rainfall might score very well on evaluations, but not be useful to users for whom a 4.9 mm and 5.1 mm period of rainfall is not particularly different.

When the definition of the phenomenon of interest depends on specific parameters, the sensitivity of the phenomenon to these thresholds should be examined in both the ground truth and metrics of forecast skill, and the evaluator should confirm that small changes in the parameters do not lead to large jumps in ground truth distributions or skill.

\section*{Acknowledgments}
This work is partially supported by the Gates Foundation, 
the University of Chicago's Institute for Climate and Sustainable Growth, Founders Pledge, and the Open AI Foundation.  We are grateful for fruitful discussions with Xavier Gine, Yuki Honda, and Witold Więcek. 
\bibliographystyle{unsrt}
\bibliography{sources}

\end{document}